\documentclass[aps,pra,twocolumn,showpacs]{revtex4-1}
\usepackage{amsmath}
\usepackage{amssymb}
\usepackage{graphicx}
\usepackage{bm}
\usepackage{float}
\usepackage{setspace}
\usepackage[hyperindex,breaklinks]{hyperref}
\usepackage{subfigure}
\usepackage{rotating}
\usepackage{graphicx}
\usepackage{epstopdf}
\usepackage{epsfig}
\usepackage{amssymb}
\usepackage{mathrsfs}
\usepackage{array}

\begin{document}

\title{Observation of atom-number fluctuations in optical lattices via quantum collapse and revival dynamics}
\author{Tianwei Zhou}
\author{Kaixiang Yang}
\author{Zijie Zhu}
\author{Xudong Yu}
\author{Shifeng Yang}
\author{Wei Xiong}
\author{Xiaoji Zhou}
\author{Xuzong Chen}
\email[]{xuzongchen@pku.edu.cn}
\affiliation{School of Electronics Engineering and Computer Science, Peking University, Beijing 100871, China}
\author{Chen Li}
\author{J\"{o}rg Schmiedmayer}
\affiliation{Vienna Center for Quantum Science and Technology, Atominstitut, TU-Wien, Stadionallee 2, 1020 Vienna, Austria}
\author{Xuguang Yue}
\affiliation{State Key Laboratory of Magnetic Resonance and Atomic and Molecular Physics, Wuhan Institute of Physics and Mathematics, Chinese Academy of Sciences, Wuhan 430071, China}
\author{Yueyang Zhai}
\affiliation{Science and Technology on Inertial Laboratory, Beihang University, Beijing 100191, China}
\date{\today}
\begin{abstract}
Using the quantum collapse and revival phenomenon of a Bose--Einstein condensate in three-dimensional optical lattices, the atom number statistics on each lattice site are experimentally investigated. We observe an interaction driven time evolution of on-site number fluctuations in a constant lattice potential with the collapse and revival time ratio as the figure of merit. Through a shortcut loading procedure, we prepare a three-dimensional array of coherent states with Poissonian number fluctuations. The following dynamics clearly show the interaction effect on the evolution of the number fluctuations from Poissonian to sub-Poissonian. Our method can be used to create squeezed states which are important in precision measurement.
\end{abstract}
\maketitle

\section{Introduction}

Coherent states with well-defined amplitude and phase \cite{PhysRev.131.2766} have been widely applied in quantum optics and quantum electrodynamics, such as in the characterization of superfluidity in liquid helium-4 \cite{RN2337, RN2338} and for Bose--Einstein condensation in dilute atomic gases \cite{BEC1, BEC2}. As the
eigenstate of the annihilation operator, coherent states are formed by a superposition of different number states (Fock states). If the eigenenergy of each Fock state depends only linearly on the particle number, the states remain coherent states at all times. However, if the interactions among the underlying particles are nonlinear with respect to the particle number, a coherent state does not always remain coherent. Instead, the coherent states will exhibit a series of collapses and revivals.

An example of the quantum collapse and revival dynamics has been predicted \cite{PhysRev.140.A1051, PhysRevLett.44.1323} and observed \cite{PhysRevLett.58.353, PhysRevLett.76.1800} in the Jaynes-Cummings model for a single atom interacting with a coherent field. For a matter-wave field of a Bose--Einstein condensate (BEC), collapses and revivals have been observed for ultracold atoms trapped in a three-dimensional (3D) optical lattice \cite{RN2339, RN2340}. In the latter case, this non-equilibrium dynamics are mainly governed by the nonlinear interactions among atoms. Each atom number state acquires a nonlinear collisional phase shift and the atom number states start dephasing from each other \cite{Greiner}. This quantum phase diffusion process is related to quantum fluctuations of the atom number \cite{PhysRevLett.77.3489} and has been proposed to be important for realizing highly correlated quantum states. Recently, such collapses and revivals have been also successfully achieved and observed in quantum many-body systems containing thousands of particles \cite{Schweigler, Rauer307}.

As demonstrated in previous experiments \cite{RN2339, RN2340, PhysRevLett.106.235304, Braun3641}, adiabatically loading a BEC into the optical lattice would suppress on-site number fluctuations of the quantum states because of the interactions among atoms, which energetically disfavor such number fluctuations \cite{PhysRevLett.96.090401, PhysRevLett.104.113001}. In this paper, we directly realize a 3D array of coherent states in optical lattices through coherent control composed by standing-wave laser pulse sequences within a few tens of microseconds \cite{1367-2630-20-5-055005, LiuXX}. This shortcut loading procedure is three orders of magnitude faster compared with an adiabatic loading of optical lattices and is effective for avoiding the atom number squeezing. The initially prepared state is sufficiently characterized by a Poissonian atom number distribution with large on-site fluctuations. We then observe the time evolution of Poissonian on-site number fluctuations by holding the atoms for various amounts of time before rapidly increasing the lattice potential depth to isolate each lattice site. The following quantum collapse and revival dynamics show how the atom number statistics change, which in turn reflects the time evolution of on-site number fluctuations of ultracold atoms in optical lattices.

\section{Shortcut loading procedure}

Because the classical coherent matter-wave field of a BEC is characterized by
Glauber's coherent states, it is possible to create an array of coherent states by splitting a BEC into two or more parts. However, the splitting procedure should be non-adiabatic so that each atom remains delocalized \cite{Will}. The shortcut loading method \cite{1367-2630-20-5-055005} is an effective procedure for transferring a BEC from the ground state of a harmonic trap into the ground band of an optical lattice within several tens of microseconds. It is based on a standing-wave laser pulse sequence wherein the time duration and interval of each step are fully optimized to maximize the robustness and fidelity of the final state with respect to the target state.

We first consider the 1D lattice case for simplicity. The atoms in a light field can experience a scalar potential whose strength is proportional to the intensity of laser, when neglecting atom-atom interactions because the loading time is very short,  the single-atom Hamiltonian in the optical lattice constructed by the standing-wave laser is then given by:
\begin{equation}
\hat{H}_\mathrm{L}=-\frac{\hbar^2\nabla^2}{2m}+V_\mathrm{L}\cos^2(k_\mathrm{L} x)\,,
\end{equation}
where $\hbar$ is the reduced Planck constant, $m$ is the atom mass, $V_\mathrm{L}$ is the lattice depth and $k_\mathrm{L}$ the lattice laser wave number. The eigenstates of the Hamiltonian $\hat{H}_\mathrm{L}$ can be expressed as the Bloch states ($\hbar=1$)
\begin{equation}
\lvert n, q\rangle=u^{(n)}_q\mathrm{e}^{-\mathrm{i}qx}=\sum_{\ell}c_{n,\ell}\lvert 2\ell k_\mathrm{L}+q\rangle\,,
\end{equation}
with the band index $n$, the quasi-momentum $q$ and $\ell=0,\,\pm 1,\,\pm 2 \cdots $. For transferring a BEC with initial wave-function $\lvert\psi_\mathrm{i}\rangle$ into the ground band with $\lvert n=0, q=0\rangle$ as the target state $\lvert\psi_\mathrm{t}\rangle$, we apply a $M$-step lattice laser pulse sequence on the initial state $\lvert\psi_\mathrm{i}\rangle$ before switching on the optical lattice with the potential depth of $V_\mathrm{L}$. The state after applying the lattice laser pulse sequence is
\begin{equation}
\lvert\psi_\mathrm{f}\rangle=\prod^{\mathrm{M}}_{\mathrm{j}=1}\hat{U}_\mathrm{j}\lvert\psi_\mathrm{i}\rangle\,,
\end{equation}
where $\hat{U}_\mathrm{j}=\mathrm{e}^{-\mathrm{i}\hat{H}_\mathrm{j}t_\mathrm{j}/\hbar}$ is the evolution operator in the $j$th step, $\hat{H}_\mathrm{j}$ is the Hamiltonian corresponding to the lattice depth $V_\mathrm{j}$ and $t_\mathrm{j}$ is the pulse duration. For the target state $\lvert\psi_\mathrm{t}\rangle$, the parameters $\hat{H}_\mathrm{j}$ and $t_\mathrm{j}$ can be determined by maximizing the fidelity
\begin{equation}
F=\lvert\langle\psi_\mathrm{t}\lvert\psi_\mathrm{f}\rangle\lvert^2\,.
\end{equation}
$F$ is the efficiency of transforming initial state $\lvert\psi_\mathrm{i}\rangle$ into target state $\lvert\psi_\mathrm{t}\rangle$ after applying the lattice laser pulse sequence, and $1-F$ is the higher bands excitation rate. In practical application, we usually fix the lattice depth to $V_\mathrm{j}=V_\mathrm{L}$ or 0 and adopt a four-step lattice laser pulse sequence, where $[t_1,\,t_2,\,t_3,\,t_4]$ corresponds to the lattice depth $[V_\mathrm{L},\,0,\,V_\mathrm{L},\,0]$. By properly choosing the values of $[t_1,\,t_2,\,t_3,\,t_4]$, the BEC can be transferred into the ground band of the 1D lattice with different potential depth without exciting higher Bloch bands \cite{LiuXX}.

The shortcut loading method can be extended to the 3D cubic optical lattice because the potential is simply the sum of three superimposed 1D lattice potentials. In this case, the wave functions can be separated in the form of $\psi(\vec{r})=\psi_x(x)\psi_y(y)\psi_z(z)$ and the evolution operator can be separated as $\hat{U}\psi=\hat{U}_x\psi_x\hat{U}_y\psi_y\hat{U}_z\psi_z$. To transfer a BEC into the ground band of a cubic optical lattice, we just need to apply the same lattice laser pulse sequence in the 1D lattice case from $\hat{x}$, $\hat{y}$ and $\hat{z}$ directions simultaneously.  

In our experiment, an atomic BEC of $1\times10^5$ weakly repulsive $^{87}$Rb atoms is produced in the $\lvert F=1,\,m_F=-1\rangle$ state with no discernible thermal fraction. Here, $F$ and $m_F$ denote the total angular momentum and the magnetic quantum number, respectively, of the atom's hyperfine state. The BEC is prepared in a crossed-beam optical dipole trap with a nearly isotropic trapping frequency $2\pi\times20\ \mathrm{Hz}$ \cite{doi:10.1063/1.4982348}. A cubic optical lattice comprises three mutually orthogonal retroreflected laser beams each with a beam waist of $\sim 145\ \mathrm{\mu m}$. The light intensity of the lattice laser is stabilized by a pre-stage feedback loop to reduce the laser power fluctuation to less than $0.05\%$. A post-stage switch is used to ensure the sharp rising and falling edges of each lattice laser pulse duration are within $100\ \mathrm{ns}$, limited by the acousto-optic modulator. After the shortcut loading procedure, we instantaneously switch off the dipole trap to minimize the additional harmonic confinement of the atoms. In this configuration, up to 63,000 lattice sites are populated with an ensemble-averaged filling of $\bar{n}\approx 1.5$ per lattice site, which remains almost constant when the lattice depth is changed. 

\section{Quantum Collapse and Revival Dynamics}

A coherent state is defined to be the eigenstate of the annihilation operator $\hat{a}$ associated with the eigenvalue $\alpha$, in the form $\hat{a}\lvert\alpha\rangle=\alpha\lvert\alpha\rangle$. Because $\hat{a}$ is not Hermitian, the eigenvalue of the coherent state is a complex number $\alpha=\lvert\alpha\rvert\mathrm{e}^{\mathrm{i}\theta}$, where $\lvert\alpha\rvert$ and $\theta$ are the amplitude and phase, respectively, of the coherent state. Generally, a coherent state is given by a coherent superposition of Fock states as follows:
\begin{equation}
\lvert\alpha\rangle=\mathrm{e}^{-\lvert\alpha\rvert^2/2}\sum^{\infty}_{n=0}\frac{\alpha^n}{\sqrt{n!}}\lvert n\rangle\,.
\end{equation}
Apparently, the coherent state follows a Poissonian number distribution $P(n)=\mathrm{e}^{-\lvert\alpha\rvert^2}\lvert\alpha\rvert^{2n}/n!$. The mean atom number of the coherent state is $\langle\hat{n}\rangle=\lvert\alpha\rvert^2=\bar{n}$ with large number fluctuations as $\langle\delta\hat{n}^2\rangle=\langle\hat{n}\rangle=\bar{n}$.

For a homogeneous system with \emph{N} atoms and \emph{M} lattice sites, the initial state prepared by the shortcut loading method can be written as a product of identical single-particle Bloch waves with zero quasi-momentum
\begin{equation}
\lvert\Psi\rangle\propto\frac{1}{\sqrt{N!}}(\frac{1}{\sqrt{M}}\sum^{M}_{i=1}\hat{a}^\dag_{i})^N\lvert 0\rangle\,.
\end{equation}
In the limit of large \emph{N} and \emph{M} at fixed $N/M$, the state becomes indistinguishable in a local measurement from a coherent state \cite{1464-4266-5-2-352} and factorizes into a product of single-site states $\lvert\phi_i\rangle$, such that $\lvert\Psi\rangle\simeq\prod^{M}_{i=1}\lvert\phi_i\rangle$. The atom number distribution of $\lvert\phi_i\rangle$ is Poissonian as $\lvert\phi_i\rangle=\mathrm{e}^{-\lvert\alpha\rvert^2/2}\sum^{\infty}_{n=0}\frac{\alpha^n}{\sqrt{n!}}\lvert n\rangle$, where $\lvert\alpha\rvert^2=N/M=\bar{n}$ is the average atom number per lattice site.

Using the shortcut loading method, coherent states can be prepared in a 3D optical lattice with any potential depth. If the lattice potential is deep enough, the tunnel coupling between neighboring sites becomes very small compared with the repulsive interactions among atoms. The Hamiltonian governing the system is determined mainly by the on-site interaction energy among atoms and an energy offset to each lattice site:
\begin{equation}
\hat{H}=\frac{U}{2}\sum^{M}_{i=1}\hat{n}_i(\hat{n}_i-1)+\sum^{M}_{i=1}\epsilon_i\hat{n}_i\,,
\end{equation}
where $\hat{n}_i=\hat{a}^\dag_{i}\hat{a}_i$ counts the number of atoms on lattice site $i$, $U$ is the on-site interaction matrix element, and $\epsilon_i$ denotes the energy offset of lattice site $i$ due to an external harmonic confinement of the atoms. The interaction energy per atom pair is given by $U=4\pi\hbar^2 a_s/m\int\lvert\rho(\mathbf{r})\rvert^4\mathrm{d}^3\mathbf{r}$, where $a_s$ is the $s$-wave scattering length. Here $\int\lvert\rho(\mathbf{r})\rvert^4\mathrm{d}^3\mathbf{r}$ quantifies the overlap of the atomic densities with $n(\mathbf{r})=\lvert\rho(\mathbf{r})\rvert^2$. The energy of a Fock state $\lvert n\rangle_i$ is then approximated by the total interaction energy $E^{(i)}_n=Un(n-1)/2$ and an energy offset $\epsilon_i n$. Given that the lattice depth is nearly uniform over the atom cloud, we assume $E^{(i)}_n=E_n$ for each occupied lattice site $i$.

Now the global many-body state $\lvert\Psi\rangle$ in the optical lattice, which can be expressed as a product of single-site wave functions, evolves as
\begin{equation}
\begin{aligned}
\lvert\Psi(t)\rangle&=\prod^{M}_{i=1}\lvert\phi_i(t)\rangle\\
&=\prod^{M}_{i=1}\left(\mathrm{e}^{-\lvert\alpha\rvert^2/2}\sum^{\infty}_{n=0}\frac{\alpha^n}{\sqrt{n!}}\mathrm{e}^{-\mathrm{i}(E_n+\epsilon_i n)t/\hbar}\lvert n\rangle\right)\,.
\end{aligned}
\end{equation}
The Fock states $\lvert n\rangle$ are the eigenstates of the system and each Fock state evolves individually according to its eigenenergy. The dynamical evolution of the system can be probed by analyzing the atomic interference pattern observed after switching off the lattice potential. For a 3D array of coherent states $\lvert\phi_i\rangle$, the time-dependent momentum distribution is generally linear with respect to the ensemble average of $\lvert\langle\hat{a}\rangle(t)\rvert^2$, where $\langle\hat{a}\rangle(t)=\langle\phi_i(t)\lvert\hat{a}\rvert\phi_i(t)\rangle$,
\begin{equation}
n(\mathbf{k}, t)\propto M\bar{n}+\lvert\langle\hat{a}\rangle(t)\rvert^2\sum_{i\neq j}\mathrm{e}^{-\mathrm{i}\mathbf{k}\cdot(\mathbf{r}_j-\mathbf{r}_i)}\mathrm{e}^{-\mathrm{i}(\epsilon_j-\epsilon_i)t/\hbar}\,.
\end{equation}
For a coherent state,
\begin{equation}
\lvert\langle\hat{a}\rangle(t)\rvert^2=\bar{n}\mathrm{e}^{2\bar{n}(\cos(Ut/\hbar)-1)}\,,
\end{equation}
which shows the quantum dynamics with periodic collapses and revivals expected. Different phase shift of each atom number state result in a collapse at first, for short times ($t\ll h/U$) it can be approximated by $\lvert\langle\hat{a}\rangle(t)\rvert^2\simeq\bar{n}\mathrm{e}^{-2\bar{n}U^2t^2/\hbar^2}$, the width of this Gaussian decay defines the collapse time $t_\mathrm{col}=\hbar/(\sqrt{\bar{n}}U)$. At the revival time $t_\mathrm{rev}=h/U$, each number state acquires a collisional phase shift of an integer multiple of $2\pi$ , leading to the revival of the initial coherent state.  

\begin{figure}[!tb]
	\center
	\includegraphics[width=1\columnwidth]{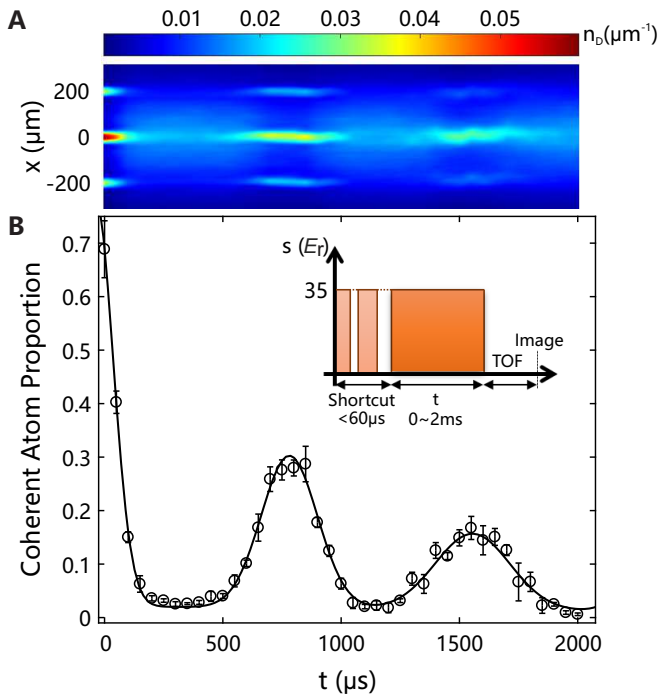}
	\caption{Quantum collapse and revival dynamics of coherent states. $\mathbf{A}$, Dynamical evolution of atomic interference pattern after the coherent states are prepared in a 3D lattice with a potential depth of $35\ E_\mathrm{r}$ and a subsequent variable hold time $t$. After the hold time, all trapping potentials are shut off and absorption images are captured after a time of flight (TOF) period of $18\ \mathrm{ms}$. The interference picture is integrated along one direction with the normalized atomic density $n_\mathrm{D}$. $\mathbf{B}$ shows the proportion of coherent atoms, i.e., the ratio of the number of atoms in the interference peaks to the total number of atoms in the interference pattern, and it is monitored over the hold time of the optical lattice. The solid line is the fit to data, comprising Gaussian curves with different widths but constant time separation, an exponential damping and a linear background. Each error bar is the standard deviation of five images. The inset shows the experimental time sequence.}
	\label{fig:fig1}
\end{figure}

By holding the atoms in the optical lattice for different lengths of time $t$ (see Fig.~\ref{fig:fig1} inset), we switch off all trapping potentials and observe the resulting atomic interference patterns after a time of flight (TOF) period of $18\ \mathrm{ms}$. An example of the dynamical evolution can be seen in Fig.~\ref{fig:fig1}($\mathbf{A}$), which shows the collapse and revival of the interference patterns over $2000\ \mathrm{\mu s}$ in a 3D lattice with a potential depth of $35\ E_\mathrm{r}$, where $E_\mathrm{r}=h^2/2m\lambda^2$ is the recoil energy and $\lambda=1064\ \mathrm{nm}$. The depth of the optical lattice is calibrated with a calibration uncertainty of less than $0.6\%$ via a method that we have used previously \cite{Zhou:18}.

We evaluate the ratio of the number of atoms in the interference peaks to the total number of atoms in the TOF images as the coherent atom proportion, which is monotonically related to $\lvert\langle\hat{a}\rangle(t)\rvert^2$ \cite{RN2339}. Figure~\ref{fig:fig1}($\mathbf{B}$) shows the proportion of coherent atoms monitored over the hold time of the optical lattice under the same experimental sequence as in Fig.~\ref{fig:fig1}($\mathbf{A}$). The first and second revivals clearly occur at the expected times, which are multiples of $t_\mathrm{rev}$. The overall damping is mainly due to the inhomogeneous dephasing that arises from the different potentials of the lattice sites. The Gaussian beam profile of the lattice laser leads to additional harmonic confinement with a trapping frequency of $60\ \mathrm{Hz}$ for a lattice potential depth of $35\ E_\mathrm{r}$. The additional harmonic confinement and the inhomogeneous density profile of the BEC both result in the change of the revival signals \cite{PhysRevA.83.043614}, and we take the initial decay time before the first revival as the collapse time $t_\mathrm{col}$.  

\section{Time Evolution of On-site Number Fluctuations}

The on-site number fluctuations are characterized by the width of the distribution of the atom number per site $\sigma$. As given in Eq. (10), at integer multiples of the revival time $t_\mathrm{rev}=h/U$, each number state acquires a collisional phase shift that is an integer multiple of $2\pi$, leading to revivals of the initial coherent state. The collapse time $t_\mathrm{col}$ depends on the width $\sigma$ of the atom number distribution \cite{PhysRevLett.78.2511, PhysRevLett.77.2158, PhysRevA.56.591, PhysRevA.55.4330}, such that $t_\mathrm{col}=\hbar/(\sigma U)$. Thus,
\begin{equation}
\frac{1}{\sigma}=2\pi\frac{t_\mathrm{col}}{t_\mathrm{rev}}\,.
\end{equation}
The on-site number fluctuations can be probed by measuring the ratio of collapse time $t_\mathrm{col}$ to revival time $t_\mathrm{rev}$. In the following experiments, we use $t_\mathrm{col}/t_\mathrm{rev}$ as the figure of merit to reflect the width of the atom number distribution.

In the first series of experiments, we use the quantum collapse and revival dynamics to verify that the initial prepared state is a coherent state with Poissonian number fluctuations. We start by preparing 3D arrays of coherent states via the shortcut loading method with a lattice depth ranging from $30\ E_\mathrm{r}$ to $50\ E_\mathrm{r}$ , where a Mott insulator should form for an adiabatic loading. As we change the depth of the optical lattice in which the initial state is prepared, the atom number distribution in each potential well is expected to remain Poissonian. 

\begin{figure}[!hbt]
	\center
	\includegraphics[width=0.95\columnwidth]{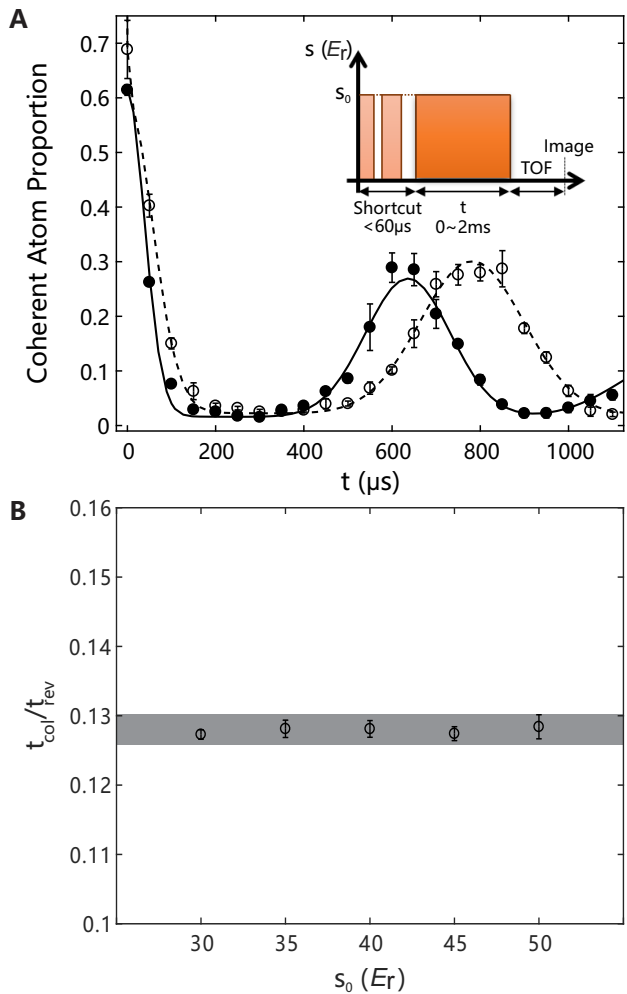}
	\caption{Realization of coherent states in optical lattices with different potential depths and the following quantum collapse and revival dynamics. $\mathbf{A}$, Initial collapse and first revival observed for lattice depths of $35\ E_\mathrm{r}$ (open circles) and $45\ E_\mathrm{r}$ (filled circles). The dashed and solid lines are fits to the data, consisting of Gaussian curves with different widths but constant time separations, corresponding to the revival time $t_\mathrm{rev}$ for each lattice depth. The collapse time $t_\mathrm{col}$ is measured as the initial decay time before the first revival, which is the first $1/e$ half width of the corresponding fitted curve. The error bars denote the standard deviation for the average taken over five images. $\mathbf{B}$, The ratio of collapse time $t_\mathrm{col}$ to revival time $t_\mathrm{rev}$ for different lattice depths. The error bars reflect the $95\%$ confidence bounds of each fit to the data. The grey shading indicates the theoretically expected value for a Poisson atom number distribution with an ensemble-averaged filling of $\bar{n}\approx 1.5$ per lattice site.}
	\label{fig:fig2}
\end{figure}

Figure~\ref{fig:fig2}($\mathbf{A}$) shows the data of the initial collapse and first revival for lattice depths of $35\ E_\mathrm{r}$ (open circles) and $45\ E_\mathrm{r}$ (filled circles). The collapse time and revival time clearly decrease for a deeper lattice depth, which is due to the increased interaction energy $U$. In our case, one expects a Poissonian atom number distribution on each lattice site, exhibiting a corresponding variance in atom number as $\sigma^2=\bar{n}$. Given an uncertainty of $\sim 15\%$ in the total atom number, the average atom number per lattice site $\bar{n}$ has an uncertainty of $\sim\pm 3\%$. According to Eq. (11), the theoretically predicted collapse and revival time ratio for a Poissonian atom number distribution is $0.128(2)\pm 0.002(1)$. This value will be much larger if there is an atom number squeezing of the coherent states. The collapse time $t_\mathrm{col}$ relative to the revival time $t_\mathrm{rev}$ for different lattice depths is shown in Fig.~\ref{fig:fig2}($\mathbf{B}$). The collapse and revival time ratio does not change with the lattice depth $s_0$, remaining consistent with the theoretically expected value for a Poisson atom number distribution in each potential well.

Next, we demonstrate how the on-site number fluctuations evolve over time in a shallow lattice. From the above discussion, the revival time $t_\mathrm{rev}$ in a deep lattice clearly depends on the interactions among atoms. Atom--atom interactions can also affect the number fluctuations of atoms in a shallow lattice. Here, we prepare two initial states with different on-site number fluctuations, namely, Poissonian and sub-Poissonian, through shortcut loading and adiabatic loading, respectively. The initial states are prepared in a 3D optical lattice with a potential depth of $10\ E_\mathrm{r}$. The lattice potential depth is in the superfluid regime so that (i) the lattice sites are not isolated but (ii) the finite repulsive interactions among atoms can lead to the emergence of the atom number squeezing \cite{PhysRevLett.96.090401, PhysRevLett.98.200405, RN2341, RN2342, RN2343, PhysRevA.84.011609}. After holding the atoms for various amounts of time $t$ at the lattice depth of $10\ E_\mathrm{r}$, we rapidly increase the lattice potential depth to $35\ E_\mathrm{r}$ in $50\ \mathrm{\mu s}$, suppressing the tunnel coupling to a negligible level and freezing the atom number distribution. This ramp rate from $10\ E_\mathrm{r}$ to $35\ E_\mathrm{r}$ is chosen to preserve the atom number statistics on each lattice site \cite{Bakr547} and ensure that all atoms remain in the ground band. 

\begin{figure}[!hbt]
	\center
	\includegraphics[width=0.95\columnwidth]{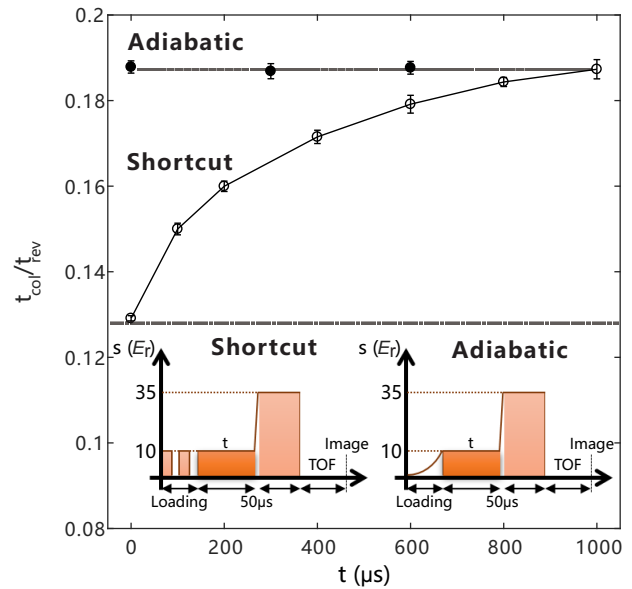}
	\caption{Time evolution of on-site number fluctuations (open circles) in a 3D lattice with a potential depth of $10\ E_\mathrm{r}$. As illustrated in Eq. (11), $t_\mathrm{col}/t_\mathrm{rev}$ is proportional to the inverse of the standard deviation of the atom number distribution. The solid line interpolates the data and the error bars reflect the $95\%$ confidence bounds of each fit to the corresponding collapse time. The filled circles are the results when we prepare the initial state by adiabatic loading a BEC into the optical lattice with a potential depth of $10\ E_\mathrm{r}$.}
	\label{fig:fig3}
\end{figure}

Corresponding to the adiabatic case in Fig.~\ref{fig:fig3}, when the 3D optical lattice potential is exponentially ramped up to the depth of $10\ E_\mathrm{r}$ in $80\ \mathrm{ms}$, with a time constant of $20\ \mathrm{ms}$, the many-body ground state of the system is a superfluid with long-range phase coherence across the lattice sites. During the ramp time, the finite repulsive interactions among atoms lead to a number squeezing of the quantum state on each lattice site. Immediately after the loading process, the atom number distribution on each lattice site is already narrowed. The on-site number fluctuations remain unchanged as sub-Poissonian over the lattice hold time $t$, i.e., the width $\sigma$ of the atom number distribution remains the same. This is consistent with the result of the ratio of collapse time $t_\mathrm{col}$ to revival time $t_\mathrm{rev}$, as can be seen from the filled circles in Fig.~\ref{fig:fig3}.

Unlike the adiabatic case, the initially prepared states of the shortcut case are coherent states with Poissonian number fluctuations. This non-adiabatic loading method avoids the number squeezing caused by the atom--atom interactions, and the many-body state after the loading process factorizes into a product of states with Poissonian number fluctuations on each lattice site. The initial Poissonian number fluctuations allow us to focus on the dynamics for narrowing the atom number distribution. As we hold the atoms in the optical lattice at $10\ E_\mathrm{r}$, the atom number statistics are still Poissonian in the beginning. The  collapse and revival time ratio is consistent with the theoretically expected value for a Poisson atom number distribution as illustrated in Fig.~\ref{fig:fig2}($\mathbf{B}$). With the extension of the lattice hold time $t$, the on-site number fluctuations tend to be suppressed because of the repulsive interactions among atoms. The on-site number fluctuations evolve gradually from Poissonian atom number statistics to sub-Poissonian atom number statistics, and the width $\sigma$ of the atom number distribution decreases with the lattice hold time $t$. The evolution of on-site number fluctuations over the lattice hold time is shown in Fig.~\ref{fig:fig3} as open circles, and the ratio of collapse time $t_\mathrm{col}$ to revival time $t_\mathrm{rev}$ increases as we continue to hold the atoms in the optical lattice. During a time of $\sim 1\ \mathrm{ms}$, the data asymptotically approach the results obtained in the adiabatic case. The final value of $\sigma$ is determined by the ratio of the on-site repulsive interaction energy $U$ to the hopping energy $J$ at the current lattice depth \cite{PhysRevA.75.013619, PhysRevLett.98.040402, PhysRevLett.104.113001}. With the observation of on-site number fluctuation evolution, the number squeezing in optical lattices can be controlled without changing the potential depth, which may lead to enhanced sensitivity for atomic interferometry \cite{Orzel2386, RevModPhys.81.1051, PhysRevLett.98.030407}. 

\section{Conclusions}

We have observed the continuous suppression of on-site number fluctuations in a constant lattice potential via quantum collapse and revival dynamics. One key merit of the measurement is the initial coherent states with Poissonian number fluctuations prepared by the shortcut loading method. The on-site number fluctuations evolve from Poissonian to sub-Poissonian atom number statistics in a timescale of $\sim 1\ \mathrm{ms}$ driven by the interactions among atoms. The ability to analyze and manipulate the atom-number fluctuations affords further potential in creating quantum states with different degrees of number squeezing.

\section*{Acknowledgments}

This work is supported by the National Key Research and Development Program of China (grant no. 2016YFA0301501) and the National Natural Science Foundation of China (grant nos. 91736208, 11504328, 61703025, 61475007, and 11334001). J.S. acknowledges support by the European Research Council through ERC-AdG: \emph{QuantumRelax}.

\bibliography{mybibfile}{}

\end{document}